\documentclass{aastex631}

\shorttitle{The doubly librating Plutinos}
\shortauthors{Malhotra and Ito}

\usepackage{amsmath}

\begin{document}

\title{The doubly librating Plutinos}

\author[0000-0002-1226-3305]{Renu Malhotra}
\affiliation{
The University of Arizona\\
1629 E University Boulevard\\
Tucson, AZ 85718, USA}

\author[0000-0002-0549-9002]{Takashi Ito}
\affiliation{
National Astronomical Observatory\\
Osawa 2--21--1\\
Mitaka, Tokyo 181--8588, Japan}



\begin{abstract}
Named for orbital kinship with Pluto, the Plutinos are a prominent group of Kuiper Belt objects whose orbital periods are in libration about the 3/2 ratio with Neptune’s.
We investigate the long term orbital dynamics of known Plutinos, with attention to the additional libration (or lack thereof) of their argument of perihelion, $g$, a well-known characteristic of Pluto's orbit. 
We show that the $g$ librators amongst the Plutinos cluster around an arc in the eccentricity--inclination parameter plane. 
This previously unreported dynamical structure is owed to a family of periodic orbits of the third kind in the restricted problem of three bodies, identified by Poincar\'e at the end of the 19th century.
Approximately sixteen percent of the currently known Plutinos exhibit $g$ librations, a far greater fraction than the ratios of the associated libration frequencies.
These results may offer new constraints for theoretical models of the dynamical history of the Plutinos and of the orbital migration history of the giant planets.
\end{abstract}



\section{Introduction\label{sec:intro}}

The Plutinos are a dynamical group of Kuiper Belt objects whose orbital period is in 3/2 resonant ratio with Neptune’s. 
Their existence and high abundance is amongst the strongest lines of evidence for the orbital migration of the giant planets during the late stages of planet formation in the ancient Solar system \citep[see, e.g., reviews by][] {Nesvorny:2018,Malhotra:2019b,Gladman:2021}. 
The defining characteristic of the Plutinos is the libration of a critical resonant angle, $\sigma$, defined as
\begin{equation}
    \sigma = 3\lambda - 2\lambda' - \varpi ,
\label{e:sigma}\end{equation}
where $\lambda$ denotes mean longitude, $\varpi$ denotes longitude of perihelion, and the primed quantities denote the parameters of Neptune. 
The critical resonant angle of every Plutino librates around a center at $\sigma=180^\circ$, with libration periods of ${\cal O}(10^4)$ years. 

Like Pluto itself, some Plutinos exhibit the additional property that their argument of perihelion, $g$, librates around one of two centers, either $g=+90^\circ$ or $g=-90^\circ$, with libration periods on the order of ${\cal O}(10^6)$ years; we call these the doubly librating Plutinos. 
(We denote the argument of perihelion with $g$, following the traditional notation of the canonical Delaunay elements of the Keplerian osculating ellipse, although $\omega$ is also commonly used in the literature for this element \citep[e.g.][]{brouwer1961}.)
The libration of $g$ is also known as the von Zeipel--Lidov--Kozai [vZLK] oscillation,
a phenomenon that has historically been studied in the context of the secular (orbit-averaged) three body problem \citep[e.g.,][note that the terms ``Kozai resonance'' or ``Lidov--Kozai resonance'' are also used in the literature]{Ito:2019,Tremaine:2023}. 
This classical model neglects the effects of mean motion resonances. 
For the case of $g$ librations in the presence of a mean motion resonance and in the presence of additional planetary perturbers, the mathematical analysis of the vZLK phenomenon requires special care in the orbit-averaging procedure \citep[e.g.,][and references therein]{Saillenfest:2020,Lei:2022,Ito:2023}.

In the present work, we examine the dynamics of the Plutinos, with attention to the doubly librating subset. 
Section \ref{sec:sims} describes the observational data of the Plutinos and our methodology for identifying the doubly librating subset. 
In Section \ref{sec:results}, we describe how the doubly librating Plutinos cluster around a specific structure in the eccentricity--inclination parameter space, 
and how this structure is related to a family of periodic orbits in the three body problem identified long ago by \cite{Poincare:1892}, specifically the ``periodic orbits of the third kind". 
Section \ref{sec:discussion} summarizes and concludes with remarks on the relative abundance of the doubly librating Plutinos and possible implications for their origins and the nature of giant planet migration.

\section{Numerical simulations\label{sec:sims}}

We retrieved the list of Plutinos identified and published in \cite{Volk:2024}.
Then, we obtained the current orbital data of these objects from JPL Horizons System (\url{https://ssd.jpl.nasa.gov/horizons/app.html#/}). 
We also obtained from the Horizons System at the same time the data for solar system parameters (masses of the Sun and the eight major planets and the orbital parameters of the planets). 
We used these data as the inputs for numerical orbit propagation for a time span of 100 myr, with the SWIFT\_RMVS3 code \citep{Levison:1994}. SWIFT\_RMVS3 is a regularized mixed variable symplectic integrator optimized for fast and accurate propagation of planetary orbits. 
The Plutinos were treated as test particles in this numerical simulation. 
The initial propagation step size was chosen as 30 days, and we recorded the orbital evolution of planets and test particles every 500 years. 

From the time series of the orbit propagation results, we identified 441 long term stable Plutinos, using the criterion that the range of the critical resonant angle, $\sigma$~modulo~360$^\circ$, over the 100 myr time span remained limited: $\mathrm{max}\{\sigma\}-\mathrm{min}\{\sigma\} < 358^\circ$. 
These identifications were verified by visual inspections of plots of $\sigma(t)$. 
The reduction in the number of surviving Plutinos from the initial sample size of 453 to 441 is due to our longer integration time span (100 myr) compared to the 10 myr time span used in \cite{Volk:2024}. 
Even so, this sample size is significantly larger than in previous studies of the orbital characteristics of known Plutinos.
For comparison, \cite{Lawler:2013} and \cite{Volk:2016} studied Plutino samples of less than 25, \cite{Balaji:2023} worked with a Plutino sample of 85.

Within this sample of 441 long term stable Plutinos we found 69 cases of $g$ librators, that is a $\sim16\%$ fraction. 
A few words are in order regarding how we identified the $g$ librators. 
Recall that $g$ is defined as the angular measure from the longitude of ascending node to the direction of pericenter of the orbit. 
The longitude of ascending node depends upon the choice of reference plane. 
JPL Horizons adopts the J2000 ecliptic as the reference plane, a practical choice for observational solar system astronomy. 
However, on secular timescales, the orbital planes of the planets precess around the plane normal to the total angular momentum vector of the solar system, the so-called invariable plane \citep{Murray:1999SSD,souami2012}.
Considering that the giant planets are the major perturbers defining the long term dynamics of the Plutinos, we adopted this plane as the reference plane for all angular orbital elements. 
We computed all orbital elements relative to the solar system barycenter.
We looked for cases in which the time series of a Plutino's $g$ showed persistent librations about a possible mean value. 
Guidance from the classical (non-resonant) vZLK theory of the circular restricted three body problem \citep[e.g.][]{Ito:2019} indicates that there are four possible centers of $g$ libration: 0, $90^\circ$, $180^\circ$ and $270^\circ$.
In the 3:2 exterior mean motion resonance, the $g$ libration is expected to be centered at these same four values, but in other resonances the $g$ librations can be centered at other values \citep{Kozai:1985, saillenfest2016, Saillenfest:2020}.
We looked for $g$ librations about all four centers, with the criterion that the range of $g$ over the entire 100 myr span remains limited to a half-circle: 
$\mathrm{max}\{g-g_c\}-\mathrm{min}\{g-g_c\} < 179^\circ$, where $g_c$ takes one of the four values (0, 90, 180, or 270 degrees).
The angular quantity $g(t)-g_c$~(modulo~$360^\circ)$) is taken to be in the range $[-180^\circ,+180^\circ]$.
Of the 69 cases of $g$ librators, none were found librating about 0 or 180 degrees. The $g$-librators were nearly equally divided between those centered at $g_c=90^\circ$ (34 objects) and $g_c=270^\circ$ (35 objects).

From the time series of the orbit propagation results, we calculated the time-averaged values of the eccentricity and of the inclination of each Plutino over 100 myr. These results are discussed in the next section.

\section{Results\label{sec:results}}

\begin{figure}
    \centering
    \vglue-0.4in
\includegraphics[width=0.98\textwidth]{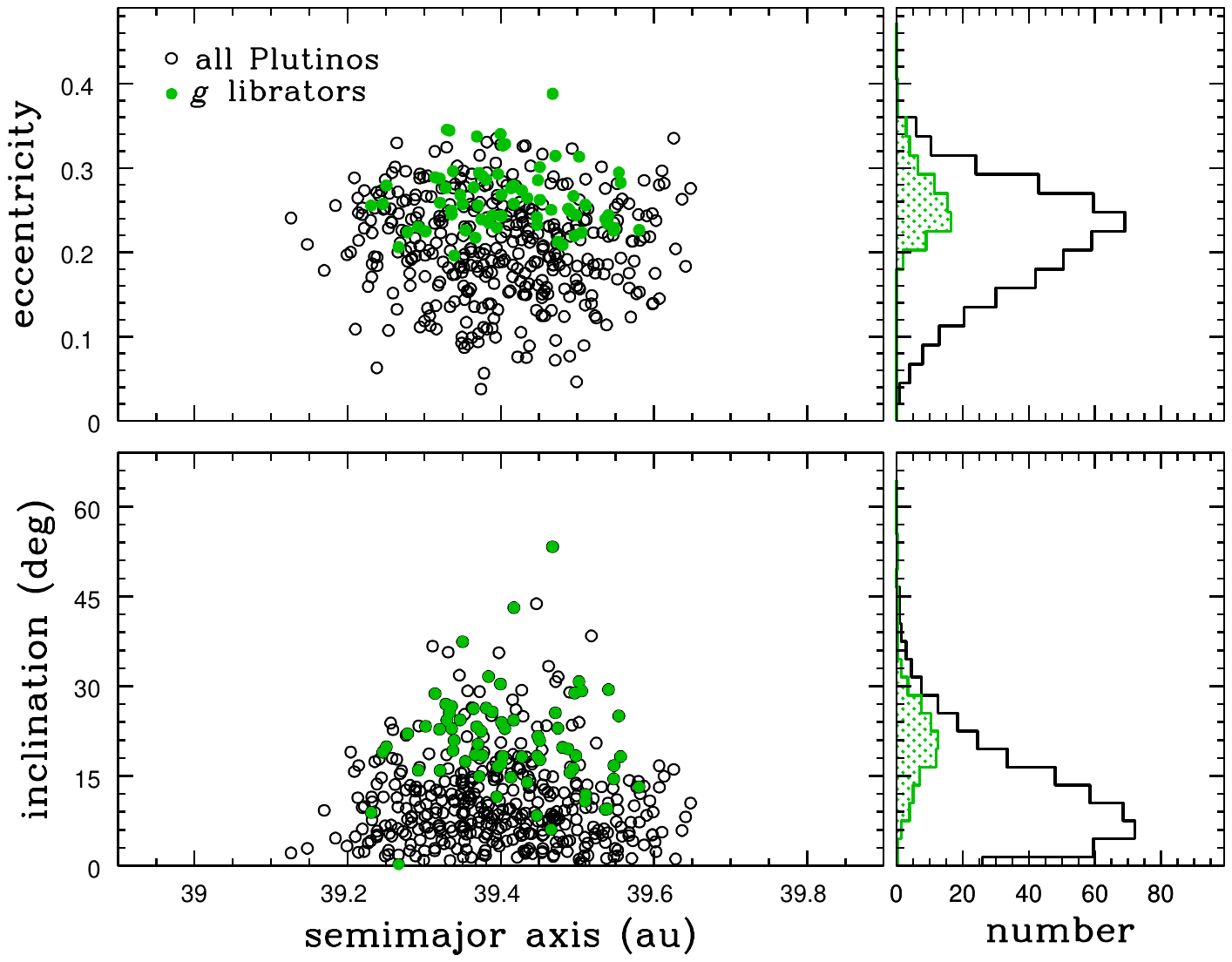}
    \vglue-0.1truein
    \caption{(Left) Scatter plots of osculating semimajor axis,  eccentricity, inclination of the 441 Plutinos in the present study.
    (Right) Histograms of the osculating eccentricities and inclinations.
    The $g$ librators are highlighted in green.} 
    \label{fig:aei0}
\end{figure}

\begin{figure}
    \centering
     \vglue-0.4in
\includegraphics[width=0.88\textwidth,angle=0]{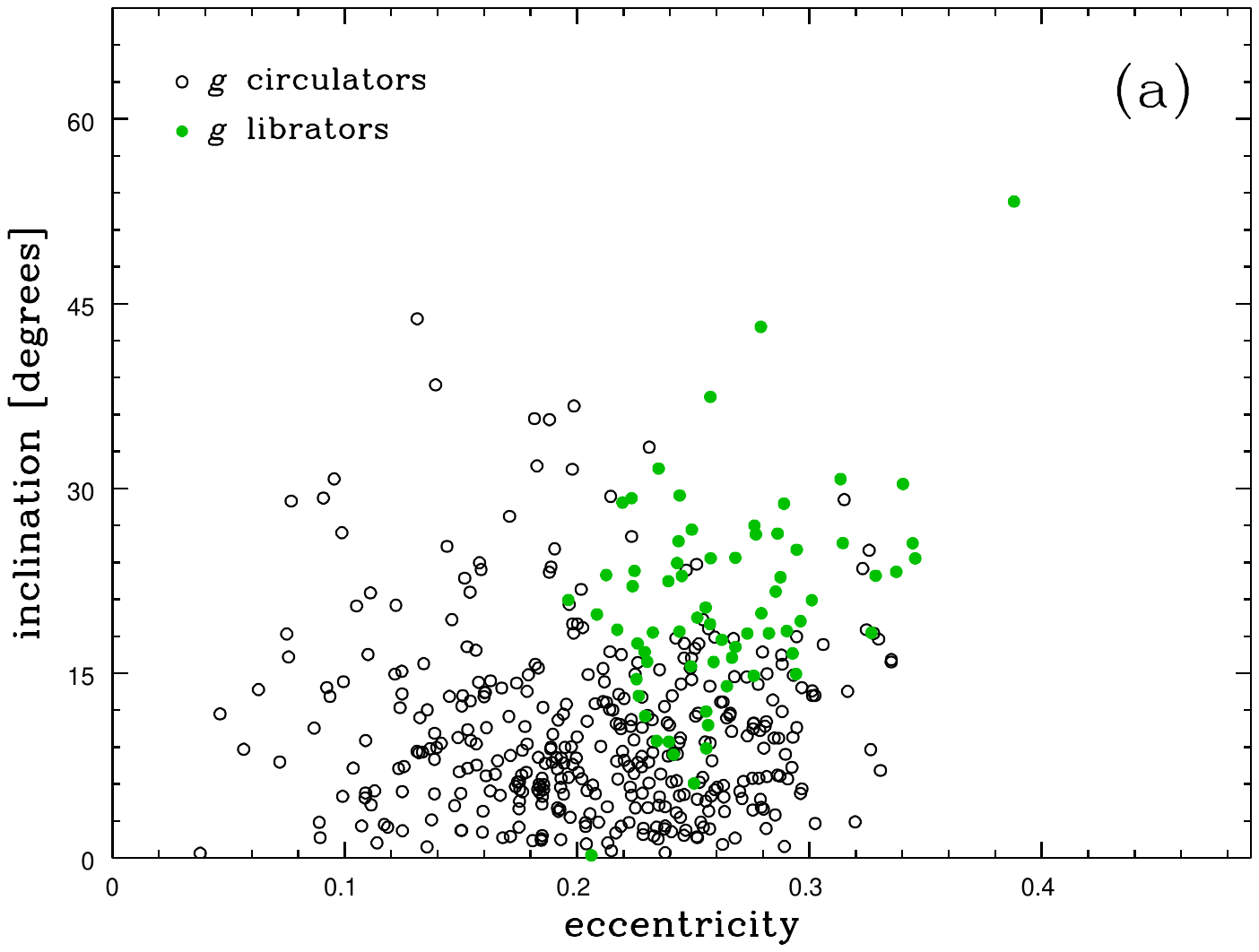}
\vglue-0.5truein
\includegraphics[width=0.88\textwidth,angle=0]{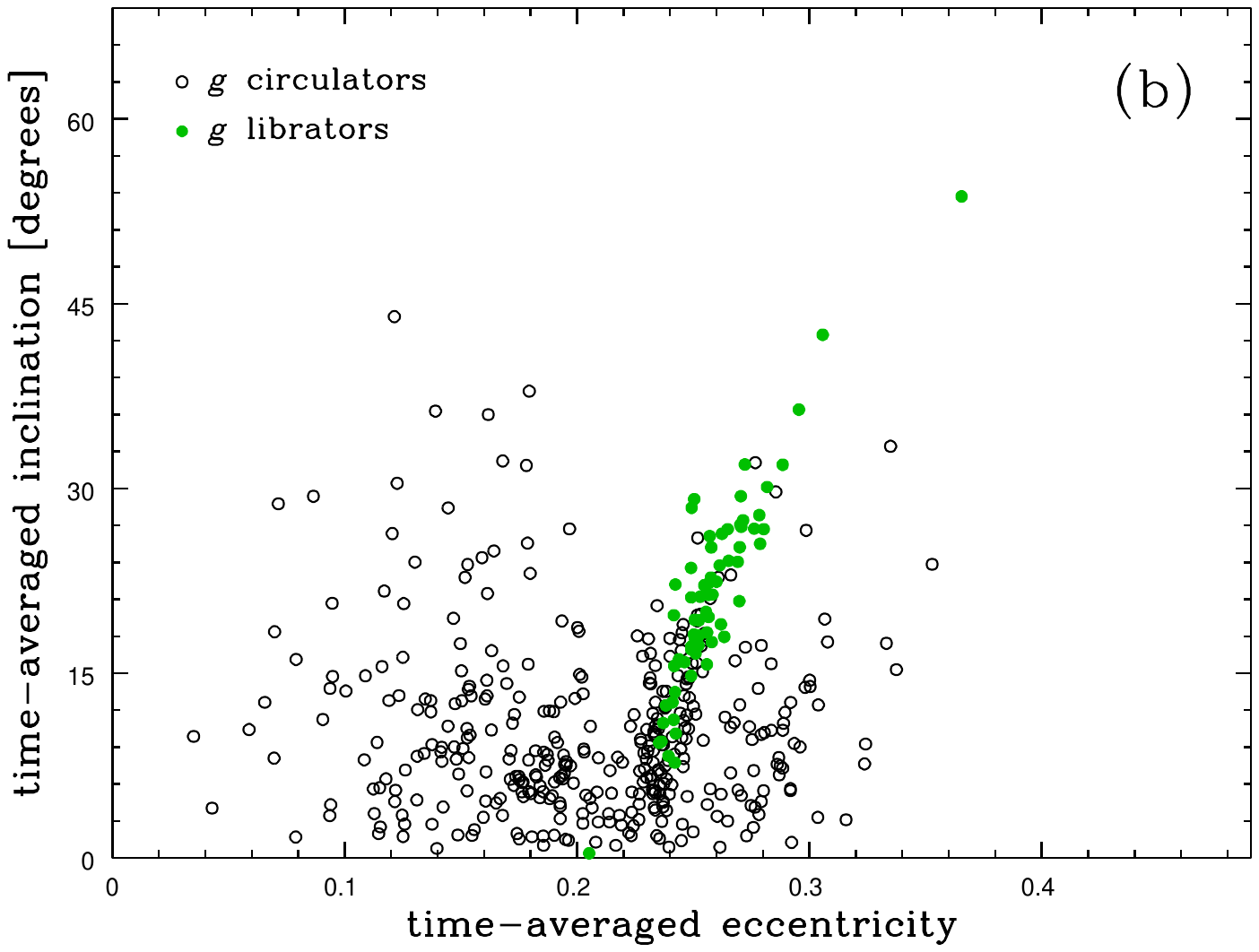}
\vglue-0.1truein
    \caption{Scatter plots of the Plutinos' eccentricities and inclinations: (a) osculating elements, (b) time-averaged elements. 
} 
    \label{fig:ei}
\end{figure}

\begin{figure}
    \centering
    \vglue-0.3truein
\includegraphics[width=0.95\textwidth,angle=0]{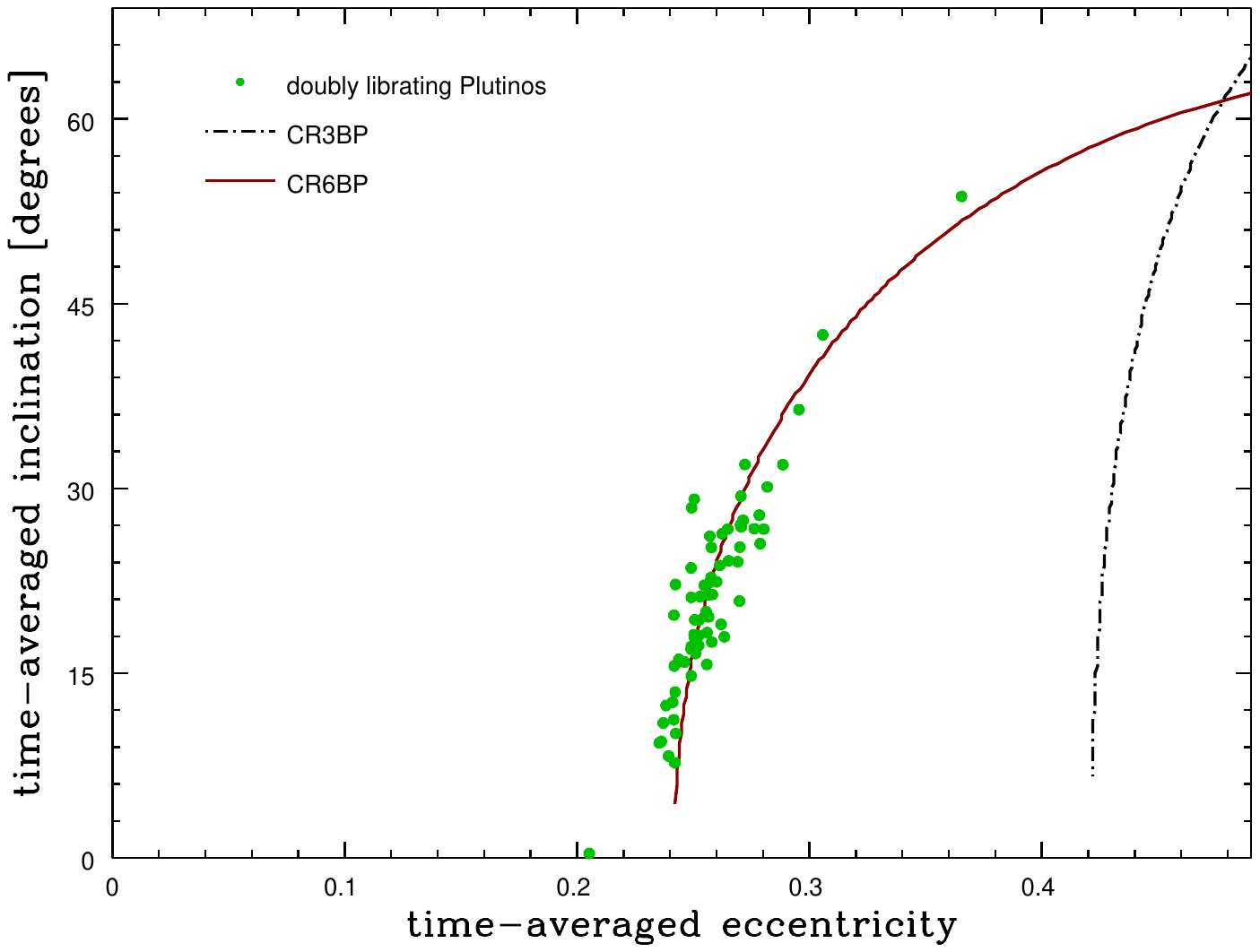}
\vglue-0.3truein
    \caption{The doubly librating Plutinos (green dots), and the locus of the $g$ libration centers of Plutinos in two models: the red curve was computed for the circular restricted six body problem [CR6BP] of the Sun, the four giant planets and a (massless) Plutino whereas the black dot-dash curve was computed for the circular restricted three body problem [CR3BP] of the Sun, Neptune and a (massless) Plutino. For both models we averaged over the mean longitudes, while taking account of the mean motion resonance constraints, to find the location of the minima of the disturbing function.
} 
    \label{fig:eith}
\end{figure}

Our main results are illustrated in three figures (Figs. \ref{fig:aei0}, \ref{fig:ei}, \ref{fig:eith}), and are as follows.

Fig.~\ref{fig:aei0} (left) show scatter plots of the osculating orbital elements (semimajor axis, $a$, eccentricity $e$, and inclination $i$) of the Plutinos at the initial epoch. 
In the histograms accompanying the scatter plots, it can be observed that eccentricities below about 0.1 are rare, there is a prominent peak near 0.25, and a cut-off above about 0.35. 
The inclinations are concentrated within a few degrees of the invariable plane, but there is a substantial tail of high inclinations reaching $\sim30^\circ$ and a few larger values (up to $\sim54^\circ$). 
These features of the Plutino eccentricities and inclinations have been noted in previous studies, albeit with smaller sample sizes \citep[e.g.,][]{Gomes:2003,li2014c,Nesvorny:2018,Gladman:2021}.

The eccentricity distribution declines sharply above $\sim0.35$.
It is largely sculpted by long term stability, as higher eccentricity Plutinos are not protected from destabilizing perturbation at close encounters with Uranus when they reach perihelion. 
At low eccentricities, the relative population is limited by the relative phase space volumes in Neptune's exterior 3:2 mean motion resonance, which has a narrow neck at low eccentricities in the $(a,e)$ plane \citep[e.g.,][]{Lan:2019,Balaji:2023}. 

The inclination distribution has been the subject of much commentary in the literature, particularly the difficulty of accounting for the abundance of higher inclinations with theoretical models \citep[e.g.,][]{Gladman:2021}.
Many observational surveys for Kuiper belt objects have been concentrated on fields closer to the invariable plane \citep[e.g.,][]{Bernstein:2004,Petit:2011,Bannister:2016}. 
Consequently, the higher inclination objects are likely under-represented in the discovered population, underscoring the high abundance of the inclination distribution. 

The $g$ librators amongst the Plutinos are highlighted in green as the filled points in the scatter plots in Fig.~\ref{fig:aei0} (left), and as the green histograms in Fig.~\ref{fig:aei0} (right). It can be observed that the $g$ librators tend to have higher eccentricity and higher inclination, compared with the overall population of Plutinos.

Fig.~\ref{fig:ei}(a) shows a scatter plot of the osculating eccentricities and inclinations, and Fig.~\ref{fig:ei}(b) shows a scatter plot of these time-averaged elements. 
A positive correlation between the osculating eccentricity and inclination of the $g$ librators is apparent to the eye. 
In the time-averaged elements this correlation comes into sharper focus as a clustering along an approximately hyperbolic arc. 
Some Plutinos that are not $g$ librators also hug the hyperbolic arc in Fig.~\ref{fig:ei}(b). 
An examination of the time series of $g(t)$ of these cases reveals that they exhibit intermittent $g$ librations, sometimes slipping between librating about a center at $g=+90^\circ$ and at $g=-90^\circ$. 
This is why they are not identified with our criterion that demands persistent $g$ librations over the whole 100 myr time span of the numerical simulations. 

To place the clustering of the $g$ librators in a physical context, we plot two theoretical curves in Fig.~\ref{fig:eith}, together with the green points indicating the $g$ librators amongst the observed Plutinos. 
The red curve is the locus in the (eccentricity--inclination parameter plane) of the local minima of the disturbing function for Plutinos, in an approximate model of the circular restricted six body problem [CR6BP] of the Sun, the four giant planets and a (massless) Plutino. 
For the purpose of computing this curve in this model, the giant planets' orbits were fixed as circular and co-planar. 
The orbit-averaged disturbing potential (averaged over the mean longitudes), is computed at the nominal center of the mean motion resonance, i.e., at $a/a_\mathrm{Neptune}=(3/2)^{\frac{2}{3}}$, while taking account of the constraint $\sigma=180^\circ$. 
We use numerical quadrature to carry out the orbit-averaging, for a wide range of the quantity $(1-e^2)\cos^2 i$ of a Plutino. 
This is a conserved quantity for a test particle in the averaged model of the circular restricted three body problem and is widely used to characterize the vZLK oscillation, both in the resonant and non-resonant cases \citep[e.g.][]{Kozai:1962,Morbidelli:2002Book,Tremaine:2023}.
This procedure follows established methods in the literature \citep[e.g., see Section 5 in][and references therein]{Saillenfest:2020,Ito:2023}. 
From these calculations, we identified the local minima of the disturbing potential at $g=90^\circ$. 
These minima are the same for $g=-90^\circ$ owing to the symmetry of the model with respect to the common plane of the planets.
The black dot-dashed curve is obtained similarly, but with the circular restricted three body model [CR3BP] of the Sun, Neptune and a Plutino; that is, it neglects Jupiter, Saturn and Uranus.

In the well-known idealized model of the CR3BP (two massive primaries on circular orbits and a test particle), the conditions encapsulated in ``$\sigma=180^\circ,\, g=90^\circ$" can be recognized as the conditions for periodic orbits of the third kind described in \cite{Poincare:1892}. 
Periodic orbits are characterized by the property that they retrace their path in space. 
In Poincar\'e's classification, periodic orbits of the first kind are those that are familiar today as the five Lagrange points of the planar circular restricted three body problem; 
they are fixed points in the rotating frame. 
The periodic orbits of the second kind are also found in the planar geometry: rather than fixed points, they trace closed curves in the rotating frame; they have mean motion in exact integer ratio with the mean motion of the primaries. 
The periodic orbits of the third kind trace three-dimensional closed curves in the rotating frame. 
We can also see that the first kind represent five cases of the coplanar 1:1 mean motion resonance whereas the second kind represent coplanar mean motion resonances of all other integer ratios. 
The third kind encompass mean motion resonances of inclined orbits with the added constraint of a stationary argument of pericenter. 
\cite{Jefferys:1966c} and \cite{Jefferys:1972} computed the trace in the eccentricity--inclination plane of such families of periodic orbits of the three-dimensional circular restricted three body problem, for many mean motion resonant ratios; 
in particular, the curve labeled ``4" in Figure 10 in \cite{Jefferys:1972} is the trace in the eccentricity--inclination plane of the periodic orbit of the third kind for the exterior 3:2 mean motion resonance at $g=\pm90^\circ$. 
Our independent computation of this curve is the black dot-dashed curve in our Fig.~\ref{fig:eith}. 
We observe that the red curve in our Fig.~\ref{fig:eith} (which is the locus of the minima of the disturbing function for the Plutinos in the averaged circular restricted six body model) resembles the black dot-dashed curve, but it is displaced towards lower eccentricities. 
With numerical experimentation, we verified that the displacement toward lower eccentricities arises from the orbit-averaged effects of the inner three giant planets, Jupiter, Saturn and Uranus.

In our computations for the red curve, we did not find any local minima of the averaged disturbing function for inclinations below $\sim4.36^\circ$ and eccentricities below 0.242. 
However, we note one exceptional case of a doubly librating Plutino below these values and distinctly away from the red curve in Fig.~\ref{fig:eith}. 
This object, 2015 VR164, has time-averaged inclination and eccentricty of $\sim0.22$~degrees and $\sim0.206$, respectively. 
Because its orbit plane nearly coincides with the reference (invariable) plane, we suspect that its $g$ libration is not owed to dynamics but to a kinematic effect related to the reference plane. 
We leave detailed analysis of this case to a future investigation.

\section{Summary and Discussion}\label{sec:discussion}

We summarize our work as follows.
\begin{enumerate}
    \item We have shown that the $g$ librators amongst the Plutinos cluster along an approximately hyperbolic arc in the time-averaged eccentricity--inclination plane (Fig.~\ref{fig:ei}b).
    \item We identified this structure with a local minimum of the orbit-averaged disturbing function for Plutinos, and also with a family of periodic orbits of the third kind in the circular restricted six-body problem of the Sun, the four giant planets and a (massless) Plutino. 
    We showed that these periodic orbits are related to the family of the same name in the circular restricted three body problem, as classified by \citet[][see also \cite{Jefferys:1972}]{Poincare:1892}, but they are shifted to lower eccentricities due to the orbit-averaged effects of the additional three planets (Jupiter, Saturn and Uranus).
    \item We obtained an updated estimate -- 16\% -- for the fraction of $g$ librators amongst the Plutinos. 
\end{enumerate}

Previous estimates of the fraction of $g$ librators amongst the observed Plutinos have reported values in the range 8\%--33\%, based on sample sizes of 6--85 Plutinos \citep{Nesvorny:2000a,Petit:2011,Volk:2016,Balaji:2023}.
Our estimate of a 16\% fraction is based on a much larger sample size of 441 Plutinos, and it is generally consistent with the previous results.
Simulations of observational survey biases suggest that the true fraction is likely 20\%--40\% higher, but is difficult to ascertain for the composite of many surveys that have contributed to the current catalog of all known Kuiper belt objects in the Minor Planet Center's catalog \citep{Lawler:2013}.

The prominent clustering of the doubly librating Plutinos in the time-averaged eccentricity--inclination plane (Fig.~\ref{fig:ei}b) suggests a local over-density. 
We ask if this is illusory. 
To answer this question, we must remember that for the three spatial degrees of freedom of a Plutino, its phase space is of six dimensions, so we must ask about the relative density in the full phase space.
One can imagine that an approximate measure of the relative density would be obtained by considering the phase space volume for $g$ librations relative to the phase space volume for $\sigma$ librations (Eq.~\ref{e:sigma}): 
if the doubly librating Plutinos were not over- or under-populated, then their population as a fraction of all Plutinos would be in the proportion of the phase space volume of the $g$ libration zone within that of the $\sigma$ libration zone. 
There is not a straightforward way of estimating these relative volumes without carrying out a very large number of numerical experiments and N-body simulations, because the shapes of these volumes are complex and include dynamical chaos effects \citep[e.g.,][]{Wiggins:1990}. 
Such complexity exists even in the simplified models, such as the CR3BP and the CR6BP, that we employed in computing the two theoretical curves in Fig.~\ref{fig:eith} for the $g$ libration centers. 
We proceed to make a rough estimate as follows. 
In the reduced one-degree-of-freedom pendulum model for resonances, the phase space volume of the libration zone is proportional to the frequency of small amplitude librations, hence inversely proportional to the libration period \citep[][section 6.1]{Tremaine:2023}.
We then make the ansatz that the phase space volume for the $g$ librations as a fraction of the phase space volume for the $\sigma$ librations is the inverse of the ratio of their respective libration periods.
In our simulations, we found that the libration period of $\sigma$ is typically $\sim20$ kyr and the libration period of $g$ is typically $\sim4$ myr, similar to those periods for Pluto itself \citep[e.g.,][]{Malhotra:1997}. 
Thus, if the phase space of the Plutinos were populated randomly and uniformly, one would expect the fraction of $g$ librators to be approximately 0.005. 
The observed fraction is larger by a factor $\sim30$, indicating that the $g$ librators are more abundant than might be expected from a uniform random distribution within the mean motion resonance.
We emphasize that this estimate of the over-density should be verified more rigorously in a future study.

There are two effects in the dynamics of Plutinos that may have contributed to populating their $g$ libration zone and producing the clustering in the time-averaged eccentricity--inclination plane.
\begin{enumerate}
\item[(i)] Long term dynamical stability is favored for the $g$ librators because their perihelion occurs well away from the invariable plane, thereby reducing perturbations from the giant planets. 
Consequently, long term chaotic diffusion out of the mean motion resonance, which affects all Plutinos, may be more effective in reducing the population of the $g$ circulators relative to the $g$ librators. 
This effect has been noted in a few previous studies of the dynamical stability of Plutinos on gigayear timescales \citep[e.g.][]{Kuchner:2002,Tiscareno:2009}.
\item[(ii)] The ancient period of orbital migration of the giant planets may have favored capture of the $g$ librators during the later stages when the migration would have slowed, enhancing capture into weaker resonances (those with longer libration periods). 
This effect has been noted in numerical simulation studies of giant planet migration models \citep[e.g.,][]{Chiang:2002,Nesvorny:2016,Pike:2017,Balaji:2023}.
\end{enumerate}

It is worth mention that additional clusterings may exist in the parameter space of time-averaged elements, such as the apparent cluster of about 40 Plutinos ($g$ circulators, not librators) centered at $e\approx0.18$, $i\approx6^\circ$ in Fig.~\ref{fig:ei}b. 
It would be interesting to assess the statistical significance of such clusterings in the future; 
if real, they may be indicative of collisional families analogous to those found in the main asteroid belt.
Indeed, the primary context in which time-averaged elements have long been used is in the identification of collisional families of asteroids \citep[e.g.,][]{Milani:2014}. 
We emphasize that in the present work, the cluster around the hyperbolic arc in Fig.~\ref{fig:ei}b is owed not to a collisional family but to the resonant potential function whose local minima support a family of periodic orbits of the third kind and the associated $g$ librations.

In addition, mutual gravitational perturbations and collisions with other Kuiper belt objects have undoubtedly also contributed to the details of the distribution of Plutinos within Neptune's exterior 3:2 mean motion resonance. 
Further studies would help assess the relative importance of these different dynamical effects. 
Future work could also investigate whether the $g$ librators and the $g$ circulators have systematic differences in their physical properties (such as their sizes, shapes and colors), to further constrain their origins. 
The observed abundance of the doubly librating Plutinos may offer new constraints on their provenance and dynamical history and on the nature of the orbital migration history of the giant planets.

\clearpage


\textit{Acknowledgments}
The results reported herein benefited from collaborations and/or information exchange within the program ``Alien Earths'' (supported by the National Aeronautics and Space Administration under Agreement No. 80NSSC21K0593) for NASA's Nexus for Exoplanet System Science (NExSS) research coordination network sponsored by NASA's Science Mission Directorate.
The numerical quadrature and orbit integration carried out for this study was largely performed at Center for Computational Astrophysics (CfCA), National Astronomical Observatory of Japan.
The authors acknowledge the FreeBSD Project for providing a reliable, open-source operating system that facilitated the computational tasks and data analyses required for this research.
The authors used Overleaf to provide a collaborative and efficient online LaTeX environment, which facilitated the preparation and formatting of this manuscript.
This study has used NASA's Astrophysics Data System (ADS) Bibliographic Services.

\bigskip
\textit{Data availability}
The data behind the figures is available at: 10.5281/zenodo.14635382
\url{https://doi.org/10.5281/zenodo.14635382}

%

\vspace{5mm}








\end{document}